

\input lanlmac


\newcount\figno
\figno=0
\def\fig#1#2#3{
\par\begingroup\parindent=0pt\leftskip=1cm\rightskip=1cm\parindent=0pt
\baselineskip=11pt
\global\advance\figno by 1
\midinsert
\epsfxsize=#3
\centerline{\epsfbox{#2}}
\vskip 12pt
\centerline{{\bf Fig. \the\figno:} #1}\par
\endinsert\endgroup\par
}
\def\figlabel#1{\xdef#1{\the\figno}}


\def\cob{\delta}
\def\ep{\epsilon}

\def\hf{{1\over 2}}

\def\R{{\bf R}}
\def\o{\over}
\def\til#1{\widetilde{#1}}

\def\bra{\langle}
\def\ket{\rangle}
\def\lf{\left}
\def\ri{\right}
\def\riya{\rightarrow}

\def\la{\lambda}

\def\h#1{\widehat{#1}}

\def\bt{\beta}

\def\al{\alpha}

\def\dag{\dagger}
\def\rt#1{\sqrt{#1}}
\def\st{\star}
\def\stb{\star_{b_0}}

\def\Ker{{\rm Ker}}
\def\Im{{\rm Im}}

\def\sitarel#1#2{\mathrel{\mathop{\kern0pt #1}\limits_{#2}}}

\def\A{{\cal A}}
\def\Ab{{\cal A}_{b_0}}
\def\AQ{{\cal A}_{Q}}
\def\tV{\widetilde{V}}
\def\tv{\widetilde{v}}
\def\tX{\widetilde{X}}
\def\tY{\widetilde{Y}}
\def\tZ{\widetilde{Z}}
\def\tT{\widetilde{T}}
\def\tb{\widetilde{b}}
\def\tp{\widetilde{+}}
\def\tL{\widetilde{L}}
\def\tR{\widetilde{R}}

\def\np#1#2#3{{ Nucl. Phys.} {\bf B#1} (#2) #3}

\def\plo#1#2#3{{ Phys. Lett.} {\bf #1B} (#2) #3}

\lref\trape{
D.~Gaiotto, L.~Rastelli, A.~Sen and B.~Zwiebach,
``Ghost Structure and Closed Strings in Vacuum String Field Theory,''
[arXiv:hep-th/0111129].
}

\lref\Witten{
E. Witten,
``Noncommutative Geometry And String Field Theory,''
Nucl.\ Phys.\  {\bf B268} (1986) 253.
}
\lref\HorowitzYZ{
G.~T.~Horowitz and A.~Strominger,
``Translations As Inner Derivations And Associativity Anomalies In Open
String Field Theory,''
Phys.\ Lett.\  {\bf B185} (1987) 45.
}
\lref\LPP{
A. LeClair, M. E. Peskin and C. R. Preitschopf,
``String Field Theory on the Conformal Plane. 1. Kinematical Principles,''
Nucl. Phys. {\bf B317} (1989) 411\semi
``String Field Theory on the Conformal Plane. 2. Generalized Gluing,''
Nucl. Phys. {\bf B317} (1989) 464.
}
\lref\Samuel{
S.~Samuel,
``The Physical and Ghost Vertices in Witten's String Field Theory,''
\plo{181}{1986}{255}.
}
\lref\GrossJ{
D. J. Gross and A. Jevicki,
``Operator Formulation of Interacting String Field Theory,''
\np{283}{1987}{1};
``Operator Formulation of Interacting String Field Theory (II),''
\np{287}{1987}{225}.
}
\lref\CST{
E.~Cremmer, A.~Schwimmer and C.~Thorn,
``The Vertex Function in Witten's Formulation of String Field Theory,''
\plo{179}{1986}{57}.
}
\lref\ItohWM{
K.~Itoh, K.~Ogawa and K.~Suehiro,
``BRS Invariance of Witten's Type Vertex,''
Nucl.\ Phys.\  {\bf B289} (1987) 127.
}

\lref\RZ{
L. Rastelli and B. Zwiebach,
``Tachyon potentials, star products and universality,''
JHEP {\bf 0109}, 038 (2001)
[arXiv:hep-th/0006240].
}

\lref\VSFT{
L.~Rastelli, A.~Sen and B.~Zwiebach,
``String Field Theory Around the Tachyon Vacuum,''
[arXiv:hep-th/0012251];
``Vacuum string field theory,''
[arXiv:hep-th/0106010].
}

\lref\KP{
V. A. Kostelecky and R. Potting,
``Analytical construction of a nonperturbative vacuum
for the open bosonic string,''
Phys. Rev. {\bf D63} (2001) 046007 [arXiv:hep-th/0008252].
}

\lref\Ohmori{
K.~Ohmori,
``A review on tachyon condensation in open string field theories,''
[arXiv:hep-th/0102085].
}

\lref\RSZ{
L. Rastelli, A. Sen and B. Zwiebach,
``Classical Solutions in String Field Theory Around the Tachyon Vacuum,''
[arXiv:hep-th/0102112];
``Half-strings, Projectors, and Multiple D-branes
in Vacuum String Field Theory,''
[arXiv:hep-th/0105058];
``Boundary CFT construction of D-branes in vacuum string field theory,''
[arXiv:hep-th/0105168].
}
\lref\GT{
D. J. Gross and W. Taylor,
``Split string field theory I,'' JHEP {\bf 0108}, 009 (2001),
[arXiv:hep-th/0105059];
``Split string field theory. II,''
JHEP {\bf 0108}, 010 (2001)
[arXiv:hep-th/0106036].
}
\lref\KO{
T.~Kawano and K.~Okuyama,
``Open string fields as matrices,''
JHEP {\bf 0106}, 061 (2001)
[arXiv:hep-th/0105129].
}
\lref\JD{
J.~R.~David,
``Excitations on wedge states and on the sliver,''
JHEP {\bf 0107}, 024 (2001)
[arXiv:hep-th/0105184].
}
\lref\Mu{
P.~Mukhopadhyay,
``Oscillator representation of the BCFT construction of D-branes in  vacuum string field theory,''
[arXiv:hep-th/0110136].
}
\lref\HK{
H.~Hata and T.~Kawano,
``Open string states around a classical 
solution in vacuum string field  theory,''
[arXiv:hep-th/0108150].
}
\lref\matsuo{
Y.~Matsuo,
``BCFT and sliver state,''
Phys.\ Lett.\ B {\bf 513}, 195 (2001)
[arXiv:hep-th/0105175];
``Identity projector and D-brane in string field theory,''
Phys.\ Lett.\ B {\bf 514}, 407 (2001)
[arXiv:hep-th/0106027];
``Projection operators and D-branes in purely 
cubic open string field  theory,''
Mod.\ Phys.\ Lett.\ A {\bf 16}, 1811 (2001)
[arXiv:hep-th/0107007].
}
\lref\Kishimoto{
I.~Kishimoto,
``Some properties of string field algebra,''
[arXiv:hep-th/0110124].
}
\lref\HataMoriyama{
H.~Hata and S.~Moriyama,
``Observables as Twist Anomaly in Vacuum String Field Theory,''
[arXiv:hep-th/0111034].
}
\lref\FO{
K.~Furuuchi and K.~Okuyama,
``Comma vertex and string field algebra,''
JHEP {\bf 0109}, 035 (2001)
[arXiv:hep-th/0107101].
}
\lref\Moeller{
N.~Moeller,
``Some exact results on the matter star-product 
in the half-string  formalism,''
[arXiv:hep-th/0110204].
}
\lref\moore{
G. Moore and W. Taylor,
``The singular geometry of the sliver,''
[arXiv:hep-th/0111069].
}

\Title{             
                                             \vbox{\hbox{EFI-01-51}
                                             \hbox{hep-th/0111087}}}
{\vbox{
\centerline{Siegel Gauge in Vacuum String Field Theory}
}}

\vskip .2in

\centerline{Kazumi Okuyama}

\vskip .2in

\centerline{ Enrico Fermi Institute, University of Chicago} 
\centerline{ 5640 S. Ellis Ave., Chicago IL 60637, USA}
\centerline{\tt kazumi@theory.uchicago.edu}

\vskip 3cm
\noindent

We study the star algebra of ghost sector in 
vacuum string field theory (VSFT).
We show that the star product of two states in the Siegel gauge is BRST
exact if we take the BRST charge to be the one found in \HK, 
and the BRST exact states
are nil factors in the star algebra. 
By introducing a new star product defined on the states 
in the Siegel gauge,  the equation of 
motion of VSFT is characterized as the projection condition 
with respect to this new product.
We also comment on the comma form of string vertex in the ghost sector.  

\Date{November 2001}

\vfill
\vfill

\newsec{Introduction}
Physics of tachyon condensation in open string theory
has been  studied extensively  
in the framework of string field theories (see \Ohmori\ for a review).
One of the big challenges in this direction 
is to describe the final state of tachyon 
condensation and see how closed strings come out.  
Vacuum string field theory (VSFT) \refs{\VSFT} was proposed 
as a possible candidate 
of the theory describing the end point of tachyon condensation. 
(See \refs{\GT\RSZ\KO\matsuo\JD\FO\Kishimoto\Mu\Moeller\HK
\HataMoriyama{--}\moore} for related papers.)
The action of VSFT has the same form as the one of Witten's 
bosonic open string field theory \Witten,
but its BRST charge consists of purely ghost terms and has trivial cohomology
which simply represents the absence of physical states of open strings
after tachyon condensation. 

A priori, there is no restriction on the form of BRST charge $Q$ in VSFT,
except for the requirement that $Q$ should be a derivation of 
the star product of
string fields. One possible form of $Q$ is given by
\eqn\Qvac{
Q=c_0+\sum_{n=1}^{\infty}f_n(c_{-n}+(-1)^nc_n)
}
where $f_n$ are arbitrary constants.
However, if one wants to solve the equation of motion
of VSFT in the Siegel gauge, the existence of the solution determines
the coefficients $f_n$ uniquely \HK.    
In this paper, we show that $Q$ obtained in \HK\ is not tied to the
specific form of the solution found in \HK\  but appears universally
in the star product of two states in the Siegel gauge.

The key observation is that $Q$ in \Qvac\ satisfies
\eqn\combQ{
\{Q,b_0\}=1.
}
Therefore, the star algebra $\A$ of string fields splits into
a direct sum of two parts as a vector space:
\eqn\AsumAbQ{
\A=\Ab\oplus\AQ,
}
where 
\eqn\defAbQ{
\Ab=\Ker(b_0)=\Im(b_0),\quad \AQ=\Ker (Q)=\Im (Q),
}
i.e., $\Ab$ and $\AQ$ are the space of states in the Siegel gauge and 
the space of BRST exact states, respectively.
They are related with each other by multiplying $b_0$ or $Q$ 
\eqn\AbQrel{
\Ab=b_0(\AQ),\quad \AQ=Q(\Ab).
}
The natural question here is that what is the multiplication rule
between $\Ab$ and $\AQ$ under the star product of string fields?
We will show that the star product of two states in the Siegel gauge
is a BRST exact state, in other words
\eqn\starAbAb{
\Ab\st\Ab\subset\AQ=Q(\Ab),
}
and $\AQ$ acts as zero under the star product:
\eqn\AQnil{
\AQ\st\A=\A\st\AQ=0.
}

This paper is organized as follows: In section 2
we show the properties of string field algebra \starAbAb\ and \AQnil\
by introducing a new set of variables which simplifies the string vertices.
In section 3, we define a new star product on $\Ab$ and discuss that
the solution of the equation of motion of VSFT is given by the projection
with respect to this new product.
In section 4, we show that the 3-string vertex of ghost sector 
can be written as a comma form as in the case of matter sector.
Section 5 is devoted to discussions.   

\vskip 2mm
\noindent
{\bf Note added}: Some of the results in this paper are discussed 
in \trape\ from the different viewpoint. 

\newsec{Star Product in the Ghost Sector}

\subsec{Review of the 3-string Vertex}
To study the structure of star algebra in the ghost sector, 
let us first review the properties of Neumann coefficients in the
3-string vertex \refs{\Samuel\GrossJ\CST\ItohWM{--}\LPP,\Kishimoto}.
The 3-string vertex in the ghost sector is written as
\eqn\Vthree{
|V_3\ket=\exp\lf(\sum_{r,s=1}^3c^{(r)\dag}\tV^{rs}b^{(s)\dag}
+c^{(r)\dag}\tv^{rs}b_0^{(s)}\ri)|+\ket_{123}
}
where $|+\ket$ is the oscillator vacuum of $b_n,c_n (n\geq 1)$ defined by
\eqn\vauplus{
|+\ket=c_0|-\ket,\quad |-\ket=c_1|0\ket,
}
and $|0\ket$ is the $SL(2,\R)$ vacuum. 
The dagger is defined by $c^{\dag}_n=c_{-n}$ and $b^{\dag}_n=b_{-n}$, 
and the summation over $n(\geq 1)$ is suppressed in \Vthree.
The Neumann coefficients $\tV^{rs}_{nm}$ and $\tv^{rs}_n$ have the following
twist properties:
\eqn\twisttrf{
C\tV^{rs}C=\tV^{sr},\quad C\tv^{rs}=\tv^{sr},
}
where $C_{nm}=(-1)^n\cob_{nm}$. They also satisfy the relations 
\eqn\relVv{\eqalign{
&\sum_{r=1}^3\tV^{rs}=C,\quad \sum_{t=1}^3\tV^{rt}\tV^{ts}=\cob^{rs},\cr
&\sum_{r=1}^3\tv^{rs}=0,\quad~~ \sum_{t=1}^3\tV^{rt}\tv^{ts}=-\tv^{rs}.
}}
It is convenient to introduce the notation
\eqn\tXdef{
\tX=C\tV^{rr},\quad \tY=C\tV^{r,r+1},\quad \tZ=C\tV^{r,r-1},
}
and
\eqn\tvdef{
\tv_0=\tv^{rr},\quad \tv_+=\tv^{r,r+1},\quad \tv_-=\tv^{r,r-1}.
}
In terms of these variables, the relations \relVv\ are written as
\eqn\XYZrel{
\tX+\tY+\tZ=1,\quad 
\tX\tY+\tY\tZ+\tZ\tX=0, \quad \tX^2+\tY^2+\tZ^2=1
}
and
\eqn\vpminvz{
\tv_+={\tZ\o\tX-1}\tv_0,\quad \tv_-={\tY\o\tX-1}\tv_0.
}

\subsec{Star Product of Coherent States in $\Ab$}
To know the property of star product on the space of Siegel gauge $\Ab$,
it is sufficient to compute the star product of coherent states constructed
on the oscillator vacuum $|-\ket$
\eqn\coherentstate{
|\al,\bt\ket=e^{\al Cb^{\dag}-c^{\dag}C\bt}|-\ket\in\Ab,
}
since they form an (over-complete) basis of $\Ab$. Here $\al$ and $\bt$
are grassmann parameters.
The star product of two coherent states is calculated as 
\eqn\starcohst{
|\al_1,\bt_1\ket\st|\al_2,\bt_2\ket=
\exp\lf(c^{\dag}\mu^3 +\sum_{r=1,2}\al_r\mu^r
-c^{\dag}\tV^{3r}\bt_r-\sum_{r,s=1,2}\al_r\tV^{rs}\bt_s\ri)|+\ket,
}
where 
\eqn\defmur{
\mu^r=\tV^{r3}b^{\dag}+\tv^{r3}b_0.
}

If we require this product state $|\al_1,\bt_1\ket\st|\al_2,\bt_2\ket$
to be BRST closed (which implies BRST exact), 
the coefficient $f_n$ in \Qvac\ is determined uniquely.
The BRST transform of this product state is
\eqn\BRSTofcohst{
Q\Big(|\al_1,\bt_1\ket\st|\al_2,\bt_2\ket\Big)=
\Big(c^{\dag}f-c^{\dag}\{Q,\mu^3\} 
-\sum_{r=1,2}\al_r\{Q,\mu^r\}\Big)|\al_1,\bt_1\ket\st|\al_2,\bt_2\ket. 
}
Therefore we obtain the condition for $f_n$ 
\eqn\condition{
f-\{Q,\mu^3\}=\{Q,\mu^1\}=\{Q,\mu^2\}=0,
}
which is written as
\eqn\condintvf{
(1-\tX)f-\tv_0=\tv_-+\tY f=\tv_++\tZ f=0.
}
Although this seems to be an over-determined system,
thanks to the identity \vpminvz\ 
we have a solution for $f_n$ 
\eqn\solfdoh{
f={1\o1-\tX}\tv_0.
}
This is exactly the one obtained in \refs{\HK,\Kishimoto}.
We conclude that the star product of two states in $\Ab$
gives a state in $\AQ$, i.e., $\Ab\st \Ab\subset\AQ$ 
for this particular choice of $Q$.

Note that since this $f$ is twist even
\eqn\twistf{
Cf=f,
}
$Q$ can be written as
\eqn\Qinvecnot{
Q=c_0+(c+c^{\dag})f.
}

\subsec{3-string Vertex on New Basis}
The result in the previous subsection can be clearly
understood by rewriting the 3-string vertex in new
variables. Notice that $f$ in \solfdoh\ 
appears in the relation of Neumann coefficients as  
\eqn\tvandtV{
\tv^{rs}=(\cob^{rs}-\tV^{rs})f.
}
This expression of $\tv^{rs}$ makes it clear that it is an eigenvector of
$\tV^{rs}$ with eigenvalue $-1$ (see \relVv).
Plugging this into \Vthree, $|V_3\ket$ becomes
\eqn\Vthreeintform{
|V_3\ket=\exp\lf(\sum_{r,s=1}^3c^{(r)\dag}\tV^{rs}
(b^{(s)\dag}-fb_0^{(s)})+\sum_{r=1}^3c^{(r)\dag}fb_0^{(r)}\ri)|+\ket_{123}.
}
Since the last term is diagonal in the index $r$ of 3-string, 
it can be absorbed by introducing a new vacuum $|\tp\ket$ by
\eqn\tpvacdef{
|\tp\ket=e^{c^{\dag}fb_0}|+\ket=U|+\ket.
}
Here $U$ is the operator defined by
\eqn\defU{
U=e^{(c+c^{\dag})fb_0}.
}
This is a unitary operator in the sense of both hermitian and BPZ conjugation:
\eqn\conjuU{
U^{\dag}={\rm bpz}(U)=U^{-1}.
}

This unitary transformation $U$ naturally accounts for the appearance of
$Q$ in the star product of Siegel gauge states. Actually, 
$Q$ is the $U$-transform of $c_0$
\eqn\QasczeroU{
Q=Uc_0U^{-1}=c_0+(c+c^{\dag})f.
}
Also the shift of $b_n$ by $f_nb_0$ in the first term of \Vthreeintform\
can be understood as the effect of $U$-transformation: 
\eqn\btrfU{
\tb_n\equiv Ub_nU^{-1}=b_n-f_nb_0.
}   
Under this unitary transformation, $b_0, c_n$ and $|-\ket$ remain intact
\eqn\Utrfpm{
Ub_0U^{-1}=b_0,\quad Uc_nU^{-1}=c_n,\quad U|-\ket=|-\ket.
}
One can easily see that 
$|-\ket$ and $|\tp\ket$ behave as a doublet under $b_0$ and $Q$
\eqn\doublettp{
b_0|\tp\ket=|-\ket,\quad Q|-\ket=|\tp\ket,
}
and the inner product of these states is the same as the original
vacua $|\pm\ket$:
\eqn\tpinnertp{
\bra -|\tp\ket=\bra \tp|-\ket=1,\quad \bra\tp|\tp\ket=\bra -|-\ket=0.
}
Finally,  $|V_3\ket$ in this new basis becomes 
\eqn\Vthinnew{
|V_3\ket=\exp\lf(\sum_{r,s=1}^3c^{(r)\dag}\tV^{rs}
\tb^{(r)\dag}\ri)|\tp\ket_{123}.
}

We can also rewrite the reflector $\bra R|$ in this new basis.
$\bra R|$ in the original variable is given by
\eqn\reflec{
\bra R|={}_{12}\bra +|\cob(b_0^{(1)}-b_0^{(2)})
\exp\lf(-c^{(1)}Cb^{(2)}-c^{(2)}Cb^{(1)}\ri).
}
In terms of the new vacuum $|\tp\ket$,
the zero mode part of reflector is rewritten as 
\eqn\Zintp{\eqalign{
{}_{12}\bra +|\cob(b_0^{(1)}-b_0^{(2)})&={}_{12}\bra \tp|
\exp\lf(c^{(1)}fb_0^{(1)}+c^{(2)}fb_0^{(2)}\ri)\cob(b_0^{(1)}-b_0^{(2)}) \cr
&={}_{12}\bra \tp|\cob(b_0^{(1)}-b_0^{(2)})
\exp\lf(c^{(1)}fb_0^{(2)}+c^{(2)}fb_0^{(1)}\ri).
}}
Therefore, the reflector in the new basis has the same form as the
original one: 
\eqn\refontp{
\bra R|={}_{12}\bra \tp|\cob(b_0^{(1)}-b_0^{(2)})
\exp\lf(-c^{(1)}C\tb^{(2)}-c^{(2)}C\tb^{(1)}\ri).
}

In summary, the 3-string vertex and reflector in this new basis 
have the same form as
the original one except for the absence of $b_0$ in the exponent in 
$|V_3\ket$. Combining this observation with the fact that 
$\Ab$ and $\AQ$ can be written as
\eqn\AbQFock{\eqalign{
&\Ab={\rm Span}\lf\{c^{\dag}_{n_1}\cdots c^{\dag}_{n_k}
b_{m_1}^{\dag}\cdots b_{m_l}^{\dag}|-\ket\ri\}
={\rm Span}\lf\{c^{\dag}_{n_1}\cdots c^{\dag}_{n_k}
\tb_{m_1}^{\dag}\cdots \tb_{m_l}^{\dag}|-\ket\ri\}, \cr
&\AQ={\rm Span}\lf\{c^{\dag}_{n_1}\cdots c^{\dag}_{n_k}
\tb_{m_1}^{\dag}\cdots \tb_{m_l}^{\dag}|\tp\ket\ri\}, 
}}
we arrive at the same conclusion $\Ab\st\Ab\subset\AQ$. 
Because of the inner product structure \tpinnertp\ and the absence of $b_0$
in \Vthinnew, 
all the elements in $\AQ$ 
behave as zero in the star algebra
\eqn\starQzero{
QA\st B=B\st QA=0\qquad \forall A,B.
}
For example, $c_0|I\ket$ belongs to $\AQ$ and hence decouples from 
the star product \refs{\RZ,\VSFT,\Kishimoto}. Note that the 
``identity string field $|I\ket$'' is not the identity of star product 
\Kishimoto\
since $|I\ket\st\AQ=0$.
\foot{Assuming $\bra A|B\ket=\bra I|A*B\ket$, the authors of \trape\ argued
that eq.\starQzero\ should be relaxed to $\bra C|QA\st B\ket=0$ 
for Fock space states $A,B,C$.}

\newsec{Reduced Star Product on $\Ab$}
Although the star product does not close on the space of 
Siege gauge states $\Ab$,
we can define a new product on $\Ab$ using the isomorphism
$\AQ\cong \Ab=b_0(\AQ)$.
We define a new product $\stb$ on $\Ab$ as
\eqn\defredst{\eqalign{
\stb:\Ab\times\Ab&\riya\Ab \cr
(A,B)&\mapsto b_0(A\st B)\equiv A\stb B,
}}
which we will call ``reduced star product''.
The original star product $\st$ and the reduced product $\stb$ are related by
\eqn\strstrel{
A\stb B=b_0(A\st B),\quad A\st B=Q(A\stb B).
}
The first equation is the definition of $\stb$. 
The second equation can be shown as
\eqn\strmapstshow{
Q(A\stb B)=Qb_0(A\st B)=(1-b_0Q)(A\st B)=A\st B.
}
In the last step, we used the fact $A\st B\in \AQ$.
This reduced product is (at least formally) associative
since this product is written in terms of $\tV^{rs}$ as in the original
star product. 
We can consistently extend this reduced product to the whole space by simply
setting the elements of $\AQ$ to be zero under this product:
\eqn\zeroAQ{
QA\stb B=B\stb QA=0 \quad\forall A,B.
}

Using this product $\stb$, we can define the wedge-like state 
$|n\ket_w\in\Ab$ as in \RZ 
\eqn\wedgelikedef{
|n\ket_w=\big(|-\ket\big)_{\stb}^{n-1},
} 
and the sliver-like state as the limit of wedge-like state
\eqn\slivlike{
|S_-\ket=\lim_{n\riya\infty}|n\ket_w.
}
In the same way as the original sliver state,
this sliver-like state turns out to be
a projection with respect to the reduced star product
\eqn\SstrS{
|S_-\ket=|S_-\ket\stb|S_-\ket.
}
Due to the relation \strstrel, $-|S_-\ket$ gives a solution of
the equation of motion of VSFT:
\eqn\eomQ{
Q|S_-\ket=Q(|S_-\ket\stb|S_-\ket)=|S_-\ket\st|S_-\ket.
}
In the oscillator representation, $|S_-\ket$ is given by \refs{\HK,\Kishimoto}
\eqn\osciSminus{
|S_-\ket={\cal N}_{\infty}e^{c^{\dag}C\tT b^{\dag}}|-\ket
}
where ${\cal N}_{\infty}$ is a normalization factor and $\tT$ is defined by
\eqn\deftildeT{
\tT={1\o2\tX}\Big[1+\tX-\rt{(1-\tX)(1+3\tX)}\Big].
}

In general, there is a one-to-one correspondence 
between the solution of the equation of motion of VSFT 
and the projection with respect to $\stb$ 
under the map of $\st$ and $\stb$ \strstrel. 
As a special projection, there is the identity of $\stb$  given by
\eqn\Iminus{
|I_-\ket={\cal N}_1e^{c^{\dag}Cb^{\dag}}|-\ket.
}

\newsec{Comma Vertex in the Ghost Sector}
As in the matter sector, we can rewrite the
3-string vertex of ghost sector in the comma form \refs{\KP,\FO}.
This form of vertex is useful to clarify the left-right split
structure of the string field algebra.  

Introducing a new set of oscillators which annihilate $|S_-\ket$ 
\eqn\deftu{
t={1\o\rt{1-\tT^2}}(b-C\tT b^{\dag}),\quad 
u={1\o\rt{1-\tT^2}}(c+C\tT c^{\dag}),
}
and making use of the formulas \KP
\eqn\fermiGAussnorm{\eqalign{
&e^{c^{\dag}Ab^{\dag}+c^{\dag}Mb+b^{\dag}Mc+cBb} \cr
=&\det[e^M(1-M)]e^{c^{\dag}(1-M)^{-1}Ab^{\dag}}e^{-c^{\dag}\log(1-M)b}
e^{-b^{\dag}\log(1-M)c}e^{cB(1-M)^{-1}b}, \cr
&\qquad (AM^T=MA,\quad BM=M^TB,\quad M^2=-AB), \cr
&e^{c^{\dag}A(b+\la)}=e^{c^{\dag}(e^A-1)\la}e^{c^{\dag}Ab},
}}
the 3-string vertex constructed on $|S_+\ket=c_0|S_-\ket$
is found to be
\eqn\Vthreemu{
|V_3\ket=\exp\lf(\sum_{r,s=1}^3u^{(r)\dag}\h{V}^{rs}
t^{(s)\dag}+u^{(r)\dag}\h{v}^{rs}b_0^{(s)}\ri)|S_+\ket_{123}.
}
Here the transformed Neumann coefficients are given by
\eqn\Vhatform{
C\h{V}=(1-C\tV \tT)^{-1}(C\tV-\tT)
=\lf(\matrix{0&\tL&\tR\cr \tR&0&\tL\cr \tL&\tR&0}\ri),
}
and
\eqn\hatvform{
\h{v}={\rt{1-\tT^2}\o1-\tV C\tT}\tv={1+\h{V}C\tT\o\rt{1-\tT^2}}\tv.
}
$\tL$ and $\tR$ are orthogonal projections given by
\eqn\defLR{
\tL={\tY+\tT\tZ\o(1-\tX)(1+\tT)},\quad 
\tR={\tZ+\tT\tY\o(1-\tX)(1+\tT)}.
}
The relation \hatvform\ can be written explicitly as
\eqn\hvform{\eqalign{
\h{v}_0&={1\o\rt{1-\tT^2}}(\tv_0+\tT\tR\tv_-+\tT\tL\tv_+), \cr
\h{v}_+&={1\o\rt{1-\tT^2}}(\tv_++\tT\tR\tv_0+\tT\tL\tv_-)=-\tR\h{v}_0, \cr
\h{v}_-&={1\o\rt{1-\tT^2}}(\tv_-+\tT\tR\tv_++\tT\tL\tv_0)=-\tL\h{v}_0. \cr
}}
These transformed Neumann coefficients satisfy the same relation
as the original one: 
\eqn\relhatVhv{\eqalign{
&\sum_{r=1}^3\h{V}^{rs}=C,\quad \sum_{t=1}^3\h{V}^{rt}\h{V}^{ts}=\cob^{rs}, \cr
&\sum_{r=1}^3\h{v}^{rs}=0,\quad~~ 
\sum_{t=1}^3\h{V}^{rt}\h{v}^{ts}=-\h{v}^{rs},\cr
&\quad \h{v}^{rs}=(\cob^{rs}-\h{V}^{rs})\h{v}_0.
}}
Note that the role of $f$ in the original picture is played by $\h{v}_0$
in this transformed picture.
We can use the same trick as in section2.3
to eliminate $b_0$ in the exponent
of $|V_3\ket$ by performing a unitary transformation 
\eqn\Uincomma{
U=e^{(u+u^{\dag})\h{v}_0b_0}.
}
The final expression of the comma vertex is
\eqn\Vthree{\eqalign{
|V_3\ket&=\exp\lf(\sum_{r,s=1}^3u^{(r)\dag}\h{V}^{rs}\til{t}^{(s)\dag}
\ri)|\til{S}_+\ket_{123} \cr
&=\exp\lf(\sum_{r=1}^3u^{(r)\dag}C\tL\til{t}^{(r+1)\dag}
+u^{(r)\dag}C\tR\til{t}^{(r-1)}
\ri)|\til{S}_+\ket_{123},
}}
where $|\til{S}_+\ket=U|S_+\ket$ and $\til{t}=UtU^{-1}$.
This form of vertex explains the appearance of
projections $\tL,\tR$ in the calculation of star product 
of coherent states constructed over the sliver-like state \Kishimoto.

The BRST charge Q determined from this form of vertex is
\eqn\QasUtrfccom{
Q=Uc_0U^{-1}=c_0+(u+u^{\dag})\h{v}_0=c_0+(c+c^{\dag})\rt{1+\tT\o1-\tT}\h{v}_0.
}
Therefore, $f$ and $\h{v}_0$ should be related by
\eqn\fvshvz{
f=\rt{1+\tT\o1-\tT}\h{v}_0.
}
From \hvform, one can see that this expression of $f$ agrees with 
\solfdoh.

\newsec{Discussions}
In this paper, we studied the structure of star algebra in the ghost sector
which is summarized by \starAbAb\ and \AQnil. 
However, there are some subtleties of our result 
as shown in the following examples.

First we discuss the value of classical action in VSFT. 
If we write the action of VSFT in a form
\eqn\Sint{
S=\int\hf A\st QA+{1\o3}A\st A\st A,
}
the action for the solution $A_0$ satisfying $QA_0+A_0\st A_0=0$ becomes
\eqn\Scl{
S_0=\int{1\o6}A_0\st QA_0.
}
It seems naively that this action vanishes identically 
due to the property \AQnil\ and cannot reproduce 
the tension of $D$-branes. 
This problem is related to the subtlety of the identity string field 
in the ghost sector. Since there is no identity of star product which
defines the integral $\int$, eq.\Scl\ should be understood as 
\eqn\BPZnotI{
S_0={1\o6}\bra A_0|QA_0\ket.
}
Note that the BPZ inner product between $\Ab$ and $\AQ$ does not vanish
identically.

The second example of subtleties is the gauge invariance in VSFT.
Using \strstrel, we find that the gauge transformation of
string field is BRST exact
\eqn\gaugetrf{
\cob A=Q\ep+[A,\ep]_{\st}=Q(\ep+[A,\ep]_{\stb}).
}
If we naively use the relation between $\st$ and $\stb$,
the BRST charge around a solution $A_0$ of VSFT
also becomes  $Q$-exact 
\eqn\BRSTsol{\eqalign{
Q_B\Psi&=Q\Psi+A_0\st\Psi-(-1)^{\Psi}\Psi\st A_0\cr
&=Q(\Psi+A_0\stb\Psi-(-1)^{\Psi}\Psi\stb A_0).
}}
However, the true BRST charge around the $D25$-brane background, 
i.e., the one constructed from the matter and ghost 
energy momentum tensors, does not satisfy $Q_B\Psi\in \AQ$ 
for a generic $\Psi$. 

These examples tell us that 
we should carefully treat the anomalous behavior of star algebra
associated with the infinite dimensional nature of Neumann coefficients
such as associativity anomaly \HorowitzYZ, 
twist anomaly \HataMoriyama\ and the singularity of sliver \moore,
which can be easily missed in the naive computation.

\centerline{{\bf Acknowledgments}}
I would like to thank I. Kishimoto for valuable discussions.
I am also grateful to Prof. Y. Nambu for his  continuous encouragement.

\listrefs

\end